# Optimal scheduling of island integrated energy systems considering multi-uncertainties and hydrothermal simultaneous transmission: A deep reinforcement learning approach

Yang Li [a], Fanjin Bu[a], Yuanzheng Li [b,*], Chao Long [c]

[a] Key Laboratory of Modern Power System Simulation and Control & Renewable Energy Technology, Ministry of Education (Northeast Electric Power University), Jilin, 132012, China

[b] School of Artifcial Intelligence and Automation, Huazhong University of Science and Technology, Wuhan 430074, China

[c] School of Water, Energy and Environment, Cranfield University, Cranfield MK43 0AL, UK

[*] Corresponding author. E-mail address: Yuanzheng_Li@hust.edu.cn (Y.Z. Li).

**Abstract** Multi-uncertainties from power sources and loads have brought significant challenges to the stable demand supply of various resources at islands. To address these challenges, a comprehensive scheduling framework is proposed by introducing a model-free deep reinforcement learning (DRL) approach based on modeling an island integrated energy system (IES). In response to the shortage of freshwater on islands, in addition to the introduction of seawater desalination systems, a transmission structure of "hydrothermal simultaneous transmission" (HST) is proposed. The essence of the IES scheduling problem is the optimal combination of each unit's output, which is a typical timing control problem and conforms to the Markov decision-making solution framework of deep reinforcement learning. Deep reinforcement learning adapts to various changes and timely adjusts strategies through the interaction of agents and the environment, avoiding complicated modeling and prediction of multi-uncertainties. The simulation results show that the proposed scheduling framework properly handles multi-uncertainties from power sources and loads, achieves a stable demand supply for various resources, and has better performance than other real-time scheduling methods, especially in terms of computational efficiency. In addition, the HST model constitutes an active exploration to improve the utilization efficiency of island freshwater.

## 1 Introduction

In order to realize the stable supply of multiple energy demands in the island, it becomes an inevitable choice to build an IES for the island. IES can coordinate the optimal operation of different energy subsystems in the region and realize the clean and sustainable supply of multiple energy demands [1, 2]. However, the output of renewable energy on the power generation side has great volatility and randomness, the load level on the user side presents large time differences, and the types of resource demand on the user side are diversified [3-4]. Therefore, the key to improving the economic dispatching level of an island IES is to improve the ability of the dispatch system to handle multi-uncertainties [5]. Meanwhile, DRL methods can adaptively perceive changes in the external environment and make dynamic decisions over time, which provide a new direction to explore for the optimal operation of island IES [6].

Some pioneering research work on island IES have been carried out. A wind-diesel-storage island electric power supply system in [7] was constructed to achieve a stable electric power supply to residents on isolated islands. To realize the stable supply of freshwater for residents, seawater desalination devices was introduced into island IES [8]. Considering the unique natural conditions of islands, it is an effective approach to solve the energy shortage problem by strengthening the use of renewable energy [9-10]. Strengthening the use of a variety of renewable energy, including wind energy, solar energy and biomass energy, is an effective way to alleviate the island energy shortage [11-12]. However, there is a high degree of uncertainty in the output of renewable energy [13,14]. Regarding the treatment of uncertainty, many relevant measures have been proposed. Based on multivariate adaptive regression splines (MARS), the authors in [15-16] use robust optimization to cope with data uncertainty and propose robust multivariate adaptive regression splines (RMARS), and the effectiveness of the approach has been demonstrated in the context of natural gas demand markets [17-18]. In addition, integrating the probability density function of renewable energy output [19], and modeling the randomness of renewable energy units by using scenario analysis method [20,21] are common methods to deal with the uncertainty of renewable energy outputs. Most research methods discussed above are model-driven approach which is based on accurate modeling of renewable energy equipment and accurate prediction of renewable energy output. They are suitable for day-ahead scheduling problems and are mostly limited to fixed power consumption plans, without sufficient consideration of uncertain changes in the load.

To reduce the dependence on equipment modeling and prediction information, DRL has been introduced into real-time energy management systems in recent years. In DRL, agents aim to obtain the maximum expected reward, which is consistent with the objective of dynamic economic scheduling in IESs. Authors in [22] combined a deep neural network with Q-learning and proposed a method for real-time optimization and management of a microgrid—deep Q network (DQN). The DQN in reference [23] realized the 'peak load shifting' of microgrids by managing the behaviors of batteries and e-vehicles through learning of historical electricity prices. In [24], a double-layer DQN algorithm was applied to a residential microgrid dispatching system. A DRL-based multi-period forecasting model is developed for microgrid scheduling in [25]. In [26], a reinforcement learning value iteration algorithm is introduced to energy management system, in which multiple agents are used to obtain the optimal scheduling policies in the sense of Nash equilibrium. However, these methods can only realize discrete control of microgrids. These methods can only be applied to simple and small energy systems; otherwise, they can cause large-dimension disaster problems. In addition, discretized actions are typically only sub-optimal actions, and their scheduling performance is moderate. Some pioneering studies applied policy-based DRL to achieve continuous control of microgrid energy management. In [27], the soft actor-critic (SAC) algorithm takes real-time electricity price and remaining power of the electric vehicle as the agents' exploration environment and aims to obtain the best charging strategy. Using DRL-based smart zoning of thermostatic loads management to handle various types of uncertainties can improve the utilization of renewable energy and the thermal comfort of users [28-29], and the authors further apply it to a grid-connected microgrids to demonstrate its good performance [30]. Motivated by these works, this study attempts for the first time to leverage DRL for energy management of island IESs. A distributed proximal policy optimization (DPPO) algorithm is adopted to simultaneously learn the uncertain changes from renewable energy sources and loads.

The above studies greatly promoted the optimized operation of island IESs, but there are still some aspects that deserve further discussion: (1) Existing studies considered resource demand in a relatively single form, and few studies comprehensively considered the demand from multiple resources, such as electricity, heat, and fresh water. (2) In view of the shortage of freshwater resources on islands, most studies focused on increasing freshwater production, and few studies aimed at improving freshwater utilization efficiency. (3) Regarding the treatment of the uncertainty of renewable energy, most studies focused on accurate modeling and forecasting of renewable energy output. However, modeling and forecasting become very difficult when considering multi-uncertainties from both power sources and loads simultaneously. (4) Dispatching schemes are mostly limited to fixed power generation plans, the uncertainty caused by the load is not fully considered, and it is difficult to handle emergencies promptly. Although some scheduling methods based on Q-learning realize real-time management of energy systems, the actions need to be discretized and the application scenarios are very limited.

In view of the above research gaps, we explored the following aspects:

(1) To meet the demand of various resources and realize the efficient use of renewable energy on islands, a novel island IES is modeled in this study. This model includes a combined heat and power unit (CHP), a combined water and power unit (CWP), a gas turbine (GT), a gas boiler (GB), and a wind turbine (WT). Besides, the scheduling of the island IES is transformed into a Markov decision process (MDP) under the framework of DRL to search for solutions.

(2) To alleviate the shortage of freshwater in island areas and improve the utilization rate of freshwater, in addition to the introduction of a seawater desalination system, a model of energy and matter co-transfer called "hydrothermal simultaneous transmission" (HST) was created. To the best of authors' knowledge, there is little or no consideration of this transmission structure in previous studies.

(3) To manage multi-uncertainties from power sources and loads, we propose a DPPO-based scheduling model. This is a model-free method that can adapt to various uncertain changes and avoids the modeling and forecasting of complex uncertainties. Therefore, it can be applied not only for day-ahead scheduling of fixed generation plans, but also for real-time scheduling.

(4) A simulation test based on real data from an island in North China was conducted to verify the superior performance of the proposed method. To prove the ability of this method to manage uncertainties, several emergencies were simulated.

**Table 1. Comparison of the proposed island IES scheduling method with the most relevant related studies**

| Reference | Resource demand | | | HST structure | Multi-uncertainties | | Application scenario | |
|---|---|---|---|---|---|---|---|---|
| | Electricity | Heat | Freshwater | | Source side | Load side | Day-ahead scheduling | Real-time scheduling |
| [7] | √ | × | × | × | √ | × | √ | × |
| [8] | √ | × | × | × | × | × | √ | × |
| [9] | √ | √ | × | × | × | × | √ | × |
| [10] | √ | √ | × | × | √ | √ | √ | × |
| [11] | √ | √ | × | × | √ | √ | √ | √ |
| [12] | √ | √ | √ | × | √ | × | √ | × |
| This paper | √ | √ | √ | √ | √ | √ | √ | √ |

Table 1 summarizes the main differences between the proposed island IES scheduling method and the most relevant related studies in the field. In the table, symbols "√" and "×" indicate whether a particular feature is considered or not, respectively.

## 2 Island IES

## 2.1 System structure

Fig. 1 shows the structure of an island IES which includes WTs, CHP units, CWP units, fast-response GTs, and fast-response GBs. Electricity is transmitted to the user side through a transmission line. The freshwater and thermal energy are transmitted to the user employing the HST mode through the same pipeline.

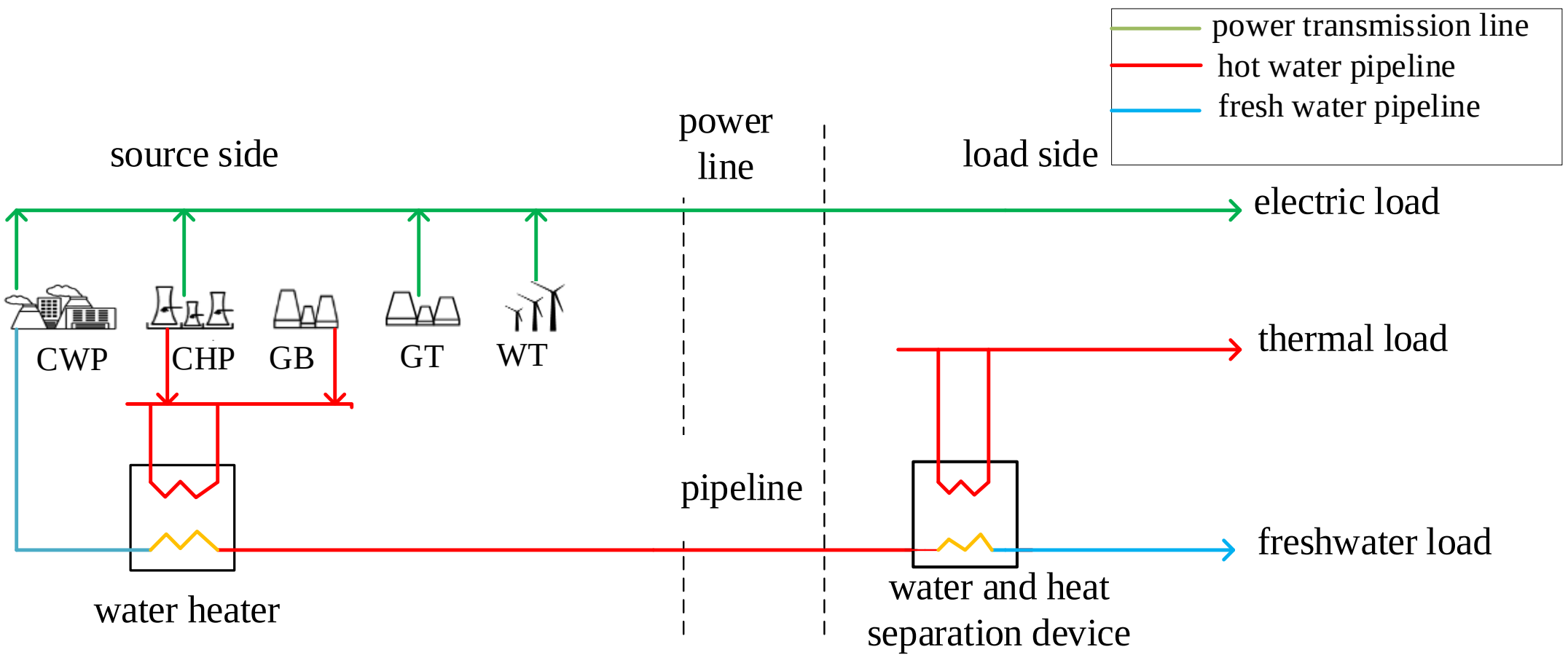

**Fig. 1. Schematic diagram of an island IES structure**

## 2.2 Combined water and power unit

### 2.2.1 Structure and operating principle of the CWP unit

The CWP unit is an efficient co-generation equipment that combines a thermal power unit and a seawater desalination system. A reverse-osmosis (RO) seawater desalination system is mainly composed of a high-pressure pump, a booster pump, an RO unit, and an energy recovery device. The requirements of RO unit for pre-water treatment are relatively high, because it is prone to contamination by suspended solids and microorganisms in the seawater, thereby becoming invalid. In addition, the permeable membrane is sensitive to the temperature of the inlet water; the higher the temperature, the faster the permeation rate of the permeable membrane. The specific relationship can be formulated as follows [31]:

$$V = V_{T_0}(1+\Delta T \cdot a)\,. \tag{1}$$

where $V$ refers to volume of water through permeable membrane per unit time, $V_{T_0}$ refers to the volume of water through permeable membrane per unit time when the temperature of the inlet water is $T_0$, $\Delta T$ is the temperature difference, and $a$ is the temperature coefficient of the permeation rate. The value of $a$ is approximately 0.02-0.03; that is, for every increase of 1°C in the inlet water temperature, the permeation rate will increase by 2%-3%. Meanwhile, the three-circuit cooling water of the thermal power unit is strictly pre-treated and has a warm temperature which is suitable as the inlet water of the seawater desalination device. Therefore, thermal unit and the seawater desalination system have strong complementary characteristics, and joint operation can improve the resource utilization efficiency. Fig. 2 shows the structure and operating principle of the CWP unit.

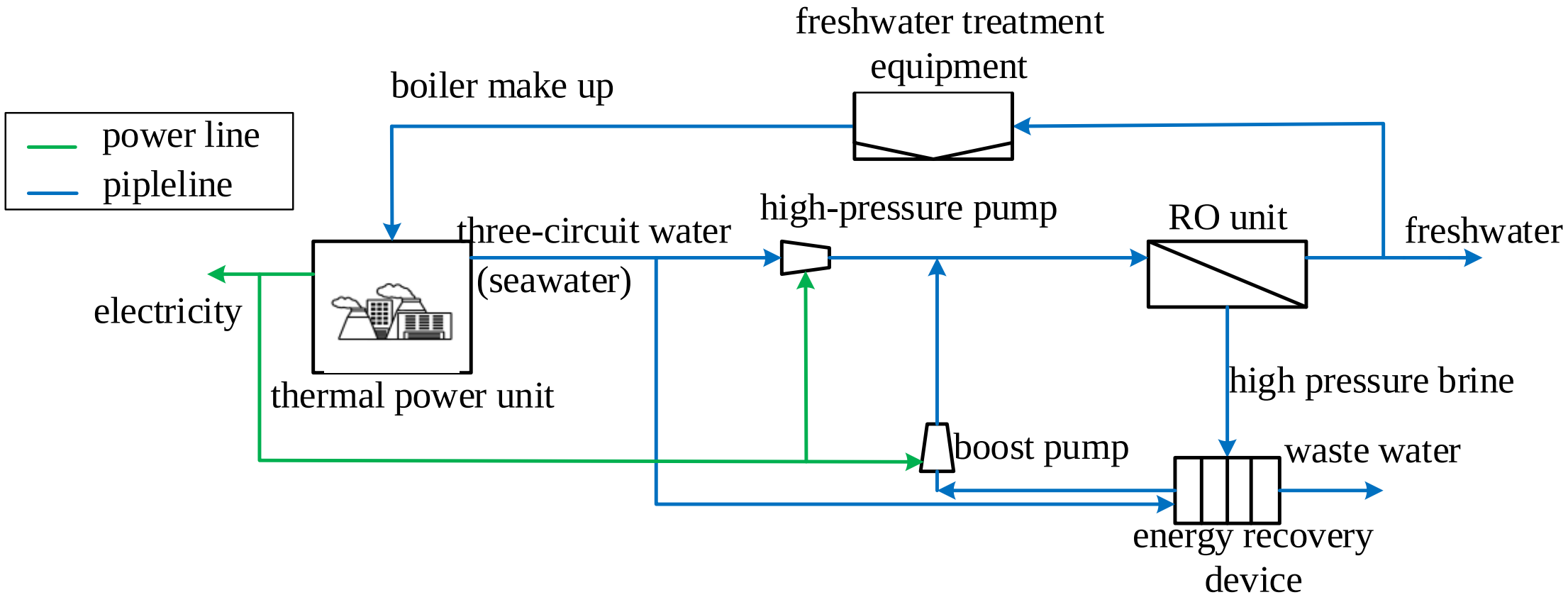

**Fig. 2. Structure diagram of a CWP unit**

The third-circuit cooling water of the thermal power unit is pressurized by a high-pressure pump and then pushed into the RO unit. Subsequently, a part of the seawater becomes freshwater through the osmotic membrane. The remaining high-salinity seawater is recovered by the energy recovery device, and after mixing with part of the third-circuit cooling water, it is sent to the RO unit again by the booster pump. The electric energy required by the high-pressure pump and booster pump is supplied by a thermal power unit, and part of the freshwater can be used as boiler compensate water for the thermal power unit after treatment. The CWP unit, jointly operated by the thermal power unit and the seawater desalination unit, realizes the co-production of freshwater and electricity through the complementary and efficient use of resources.

### 2.2.2 Energy consumption characteristics of the CWP unit

The osmotic pressure $P$ between the two sides of the permeable membrane is proportional to the seawater salinity. The specific relationship can be described by the following expression [32]:

$$P = \lambda \vartheta_t = \frac{\lambda \vartheta_0}{1-B}, \tag{2}$$

where $\lambda$ denotes the osmotic pressure coefficient, $\vartheta_t$ is the seawater salinity in the current RO unit, $\vartheta_0$ is the initial seawater concentration, and $B$ refers to the recovery rate, that is, the ratio of the volume of output freshwater to the volume of inlet seawater, which can be described by the following formula:

$$B = \frac{V_o}{V_i}, \tag{3}$$

where $V_0$ and $V_i$ are the volumes of output freshwater and inlet seawater of the RO unit, respectively. Thus, the hydroelectric conversion coefficient $Q$ that is, the electricity required to produce a unit volume of freshwater, is expressed as follows:

$$Q = \frac{1}{V}\int_0^V P dV = \lambda \vartheta_0 \frac{1}{B} \ln(\frac{1}{1-B}), \tag{4}$$

Note from this formula that there is a significant concave function relationship between the conversion coefficient $Q$ and the recovery rate $B$. Therefore, choosing the most suitable recovery rate $B$ can minimize the value of $Q$. Given that the osmotic pressure coefficient $\lambda$, seawater concentration $\vartheta_0$, and recovery rate $B$ are all fixed, the electricity $Q$ required to produce a unit volume of freshwater is also a fixed value .The electric power used by the desalination system to produce freshwater is expressed as follows:

$$p_{ro,t} = Qw_{cwp,t}. \tag{5}$$

where $p_{ro,t}$ refers to the electric power required by the desalination system, and $w_{cwp,t}$ is the freshwater production rate. The external output power of the CWP unit is defined by the following equation:

$$p_{cwp,t} = p_{tp,t} - p_{ro,t}. \tag{6}$$

where $p_{tp,t}$ is the electric power of the thermal power unit.

### 2.3 Hydrothermal simultaneous transmission model

The physical structure of the HST model is shown in Fig. 1 On the power generation side, the heat generated by the CHP unit and the GB heats the freshwater through the water heater, and then the heat-carrying freshwater is transmitted to the load side through the pipeline to realize the common transmission of material and energy. The water and heat separation device on the load side can separate the freshwater and heat energy and then supply the heat and freshwater users, respectively.

Generally speaking, the heat transfer system is mainly affected by two aspects: 1) the thermal efficiency of the heating and separation devices, and 2) the thermal attenuation caused by the heat radiation effect of the pipeline. Heating and separation devices are similar to heat exchange stations in traditional heating networks, they have high thermal efficiency. The basic structure of the core element water heater is shown in Fig. 3.

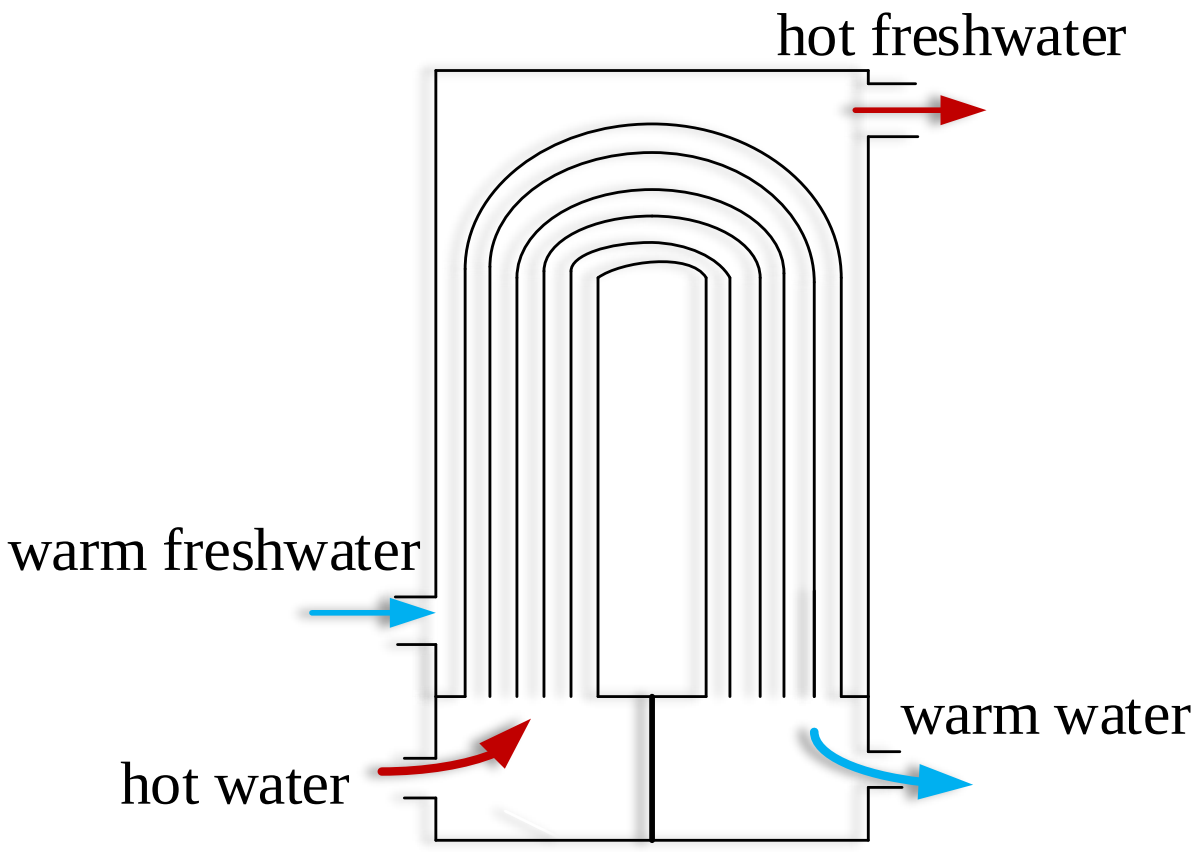

**Fig. 3. Structure of the heat exchanger**

Given that the heating and separation devices have high thermal efficiency, the heat attenuation caused by the heat radiation effect of the pipeline is mainly discussed. According to the quality-regulation mode of an HST network [33], the heat energy $H_t$ carried by the pipeline can be calculated as follows:

$$H_t = C^W G\left(T_t^{sw} - T_t^{rw}\right), \tag{7}$$

where $C^W$ refers to the specific heat capacity of water, $G$ refers to the mass of water flowing through the pipeline per unit time, $T_t^{sw}$ is the temperature of the freshwater flowing out of the heating device, and $T_t^{rw}$ is the temperature of the freshwater flowing out of the separation device. The heat loss $\Delta H_t$ of the transmission pipeline can be calculated by the following formula [33]:

$$\Delta H_t = 2\pi \frac{T_t^{sw} - T_t^{e}}{\Upsilon} L . \tag{8}$$

where $T_t^e$ represents the temperature of the outside environment, and $\Upsilon$ refers to the heat transfer coefficient between the pipeline and the outside environment. Therefore, the heat required for each scheduling period is the sum of the heat load and the transmission loss. Given that the value of $C^W$ for water is very large, it can fully meet the demand as a heat transfer carrier. The material and energy co-transmission mode of HST requires only one transmission pipeline which greatly reduces the manpower and material costs during the construction period, and saves a lot of freshwater resources during the operation period. The HST demonstration project located on the coast of Shandong Province, China, has verified the feasibility and superiority of the proposed approach after a heating season [34].

2.4 Fast-response GT and GB

Compared with steam turbine, gas turbine has the advantages of no need of preheating, fast ramp rate. It can quickly change power to deal with various uncertainties and emergencies of IES. The power change curve of gas turbine from spinning reserve state to full power operation state is shown in Fig. 4.

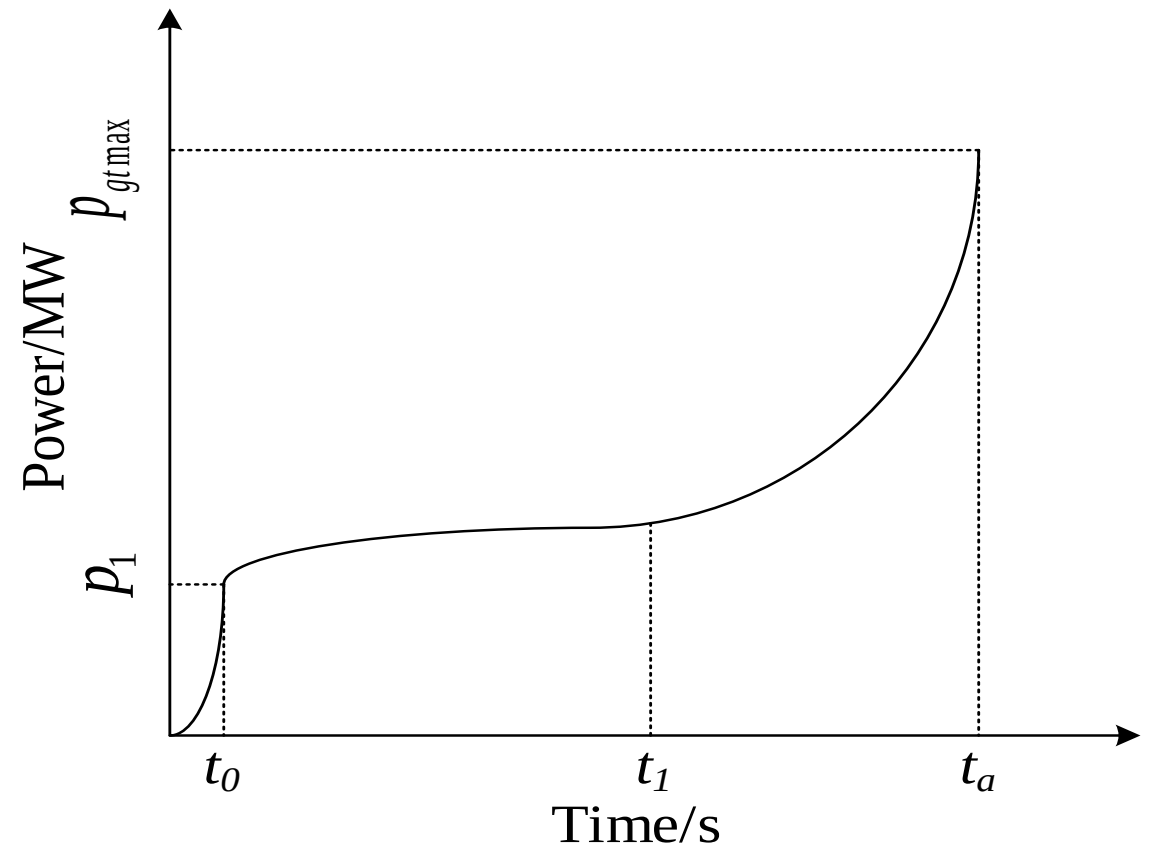

**Fig. 4. The power change curve of GT**

Generally speaking, total time $t_a$ for GT from hot reserve to full power operation will not exceed 60s. Even if it starts from cold shutdown state, the whole startup process will be within ten minutes. Therefore, the ramp rate constraints of GT are not considered in the following article. The situation of GB is similar and will not be repeated here.

## 3 Economic scheduling model of island IES

### 3.1 Objective function

In this study, the optimization objectives of the island's integrated energy system include operating costs as well as environmental costs:

$$F = \min(C_E + C_G) ,$$

where, $C_G$ is the operating cost, which is the primary energy cost in this paper; $C_E$ is the environmental cost, which is the carbon emission cost in this paper.

3.1.1 Operating cost

The specific composition of $C_G$ is given by the following formula:

$$C_G = \sum_1^{n_a} M_a\left(p_{chp}, h_{chp}\right) + \sum_1^{n_b} M_b\left(p_{cwp}, w_{cwp}\right) + \sum_1^{n_d} M_d\left(p_{gt}\right) + \sum_1^{n_e} M_e\left(h_{gb}\right), \tag{10}$$

where $M_a$, $M_b$, $M_d$, and $M_e$ denote the cost coefficients of CHP units, CWP units, GTs, and GBs; $n_a$ , $n_b$ , $n_d$ , and $n_e$ denote the number of various devices; $p_{chp}$ and $h_{chp}$ denote the electric power and thermal power of CHP units, respectively; $p_{cwp}$ and $w_{cwp}$ denote the electric power and freshwater production rate of CWP units, respectively; $p_{gt}$ refers to the electrical power of the GT; $h_{gb}$ refers to the heat power of the GB. For simplicity, the gas price and gas service time are used to calculate the operating costs; thus, the primary energy costs can be expressed as

$$M_a\left(p_{chp,} h_{chp}\right) = \rho_{gas}\left(\frac{p_{chp,t}}{\eta_a}\right)\Delta t + \rho_{gas}\left(\frac{h_{chp,t}}{\eta_a}\right)\Delta t\,, \tag{11}$$

$$M_b\left(p_{cwp,} w_{cwp}\right) = \rho_{gas}\left(\frac{p_{cwp,t}}{\eta_b}\right)\Delta t + Q\rho_{gas}\left(\frac{w_{cwp,t}}{\eta_b}\right)\Delta t\,, \tag{12}$$

$$M_d\left(p_{gt}\right) = \rho_{gas}\left(\frac{p_{gt,t}}{\eta_d}\right)\Delta t\,, \tag{13}$$

$$M_e\left(p_{gb}\right) = \rho_{gas}\left(\frac{h_{gb,t}}{\eta_e}\right)\Delta t\,. \tag{14}$$

where $\rho_{gas}$ is the price of natural gas per unit of calorific value, $\eta$ is the efficiency of each device, and $\Delta t$ refers to the gas consumption time.

3.1.2 Environmental cost

The main primary energy used in this study is natural gas, resulting in very low levels of sulfide and nitrogen compounds in the emitted gas, so only carbon emission cost is considered in the environmental cost. It can be calculated by the following formula:

$$C_G = \eta . E^{NG}(H_a + H_b + H_d + H_e)\,, \tag{15}$$

where $\eta$ is the price of CO2 emissions per unit mass, $E^{NG}$ is the carbon emissions per unit calorific value of natural gas combustion. $H_a$, $H_b$, $H_d$ and $H_e$ are respectively the calorific values of natural gas consumed by CHP units, CWP units, GTs, and GB, which are calculated by

$$H_a = \left(\frac{p_{chp,t}}{\eta_a}\right)\Delta t + \left(\frac{h_{chp,t}}{\eta_a}\right)\Delta t\,, \tag{16}$$

$$H_b = \left(\frac{p_{cwp,t}}{\eta_b}\right)\Delta t + \left(\frac{w_{cwp,t}}{\eta_b}\right)\Delta t\,, \tag{17}$$

$$H_d = \left(\frac{p_{gt,t}}{\eta_d}\right)\Delta t\,, \tag{18}$$

$$H_e = \left(\frac{h_{gb,t}}{\eta_e}\right)\Delta t\,, \tag{19}$$

### 3.2 Constraints

The constraints of island IESs scheduling include energy balance and equipment operation constraints.

1) Energy balance constraints

In period $t$, the energy balance constraints of the island IES in this study include electric power balance, thermal power balance, and freshwater supply-consumption balance, which can be respectively expressed as follows [35]:

$$p_{chp,t} + p_{cwp,t} + p_{gt,t} + p_{wt,t} = p_{load,t}\ ; \tag{20}$$

$$h_{chp,t} + h_{gb,t} = h_{load,t}\ ; \tag{21}$$

$$w_{cwp,t} = w_{load,t}\ . \tag{22}$$

where $p_{wt,t}$ is the output of the WT, $p_{load,t}$ refers to the electrical load, $h_{load,t}$ denotes the thermal load, and $w_{load,t}$ is the water load. It must be pointed out that $p_{wt,t}$, $p_{load,t}$, $h_{load,t}$ and $w_{load,t}$ are highly uncertain and dynamically changing.

2) Equipment operating constraints

For a CHP unit, the ratio of thermal power to electric power is called thermoelectric ratio $b$ :

$$b = \frac{h_{chp,t}}{p_{chp,t}}\ . \tag{23}$$

Depending on whether the thermoelectric ratio $b$ can be changed, the CHP unit can be divided into fixed thermoelectric ratio unit and variable thermoelectric ratio unit. The equipment selected in this study is a variable thermoelectric ratio unit. By controlling the electric power $p_{chp}$ and thermoelectric ratio $b$ not to exceed their operating limits, it can be ensured that the CHP unit meets its operating limits [36].

$$p_{chp\min} \le p_{chp,t} \le p_{chp\max}\ , \tag{24}$$

$$b_{\min} \le b \le b_{\max}\ . \tag{25}$$

where $p_{chp\max}$ and $p_{chp\min}$ are the upper and lower electric power limits, $b_{\max}$ and $b_{\min}$ are the upper and lower thermoelectric ratio limits, respectively. In addition, the ramp rate of CHP unit should be kept within a certain range within a unit scheduling period [37,38]:

$$-RP_{chp\max} \le RP_{tp} \le RP_{chp\max}\ . \tag{26}$$

where $RP_{chp\max}$ is upper limit of ramp rate of CHP unit.

Concerning the CWP unit, the key is to use electric energy to pressurize the seawater through the permeable membrane to complete the desalination of seawater. Therefore, the sum of the electric power required for desalination and the electric power output should not exceed the operating limit of the thermal power unit. These constraints are expressed as follows:

$$p_{tp\min} \le p_{cwp,t} + Qw_{cwp,t} \le p_{tp\max}\ , \tag{27}$$

where $p_{tp\max}$ and $p_{tp\min}$ are the upper and lower electric power limits of the thermal power unit, respectively. In addition, ramp rate of CWP unit should be kept within a certain range within a unit

scheduling period:

$$-RP_{tp\max} \le RP_{tp} \le RP_{tp\max} \,. \tag{28}$$

where $RP_{tp\max}$ is upper limit of ramp rate of CHP unit.

Concerning HST system, the temperature of the hot water in the pipeline is constrained by the allowable limits of pipeline:

$$T_{\min} \le T_t^{sw} \le T_{\max} \,, \tag{29}$$

where $T_{\min}$ and $T_{\max}$ are the maximum and minimum temperatures allowed by the pipeline, respectively. The flow of freshwater per unit time in the pipeline is also affected by the diameter of the pipeline, the operating pressure of the system and so on and kept within a certain range:

$$G_{\min} \le G \le G_{\max} \,. \tag{30}$$

where $G_{\max}$ and $G_{\min}$ are the upper and lower limits of the freshwater flow allowed through the pipeline respectively

Concerning the GT and GB, their power outputs cannot exceed the corresponding power limits. The constraints in this case are expressed as follows:

$$p_{gt\min} \le p_{gt,t} \le p_{gt\max} \,, \tag{31}$$

$$h_{gb\min} \le h_{gb,t} \le h_{gb\max} \,, \tag{32}$$

where $p_{gt\max}$ and $p_{gt\min}$ are the upper and lower electric power limits of the GT unit, respectively;

and $h_{gb\max}$ and $h_{gb\min}$ are the upper and lower thermal power limits of the GT unit, respectively.

## 4 Island IES scheduling based on DPPO

### 4.1 Theoretical basis of DPPO

In DRL, agents aim to find the best action strategy that can obtain the maximum reward by continuously interacting with the environment. The results of these interactions depend only on the current state and have nothing to do with previous states [39]; therefore, it can be described in terms of an MDP, which consists of four main elements, namely ($S$, $A$, $R$, $\pi$). The MDP is shown in Fig. 5.

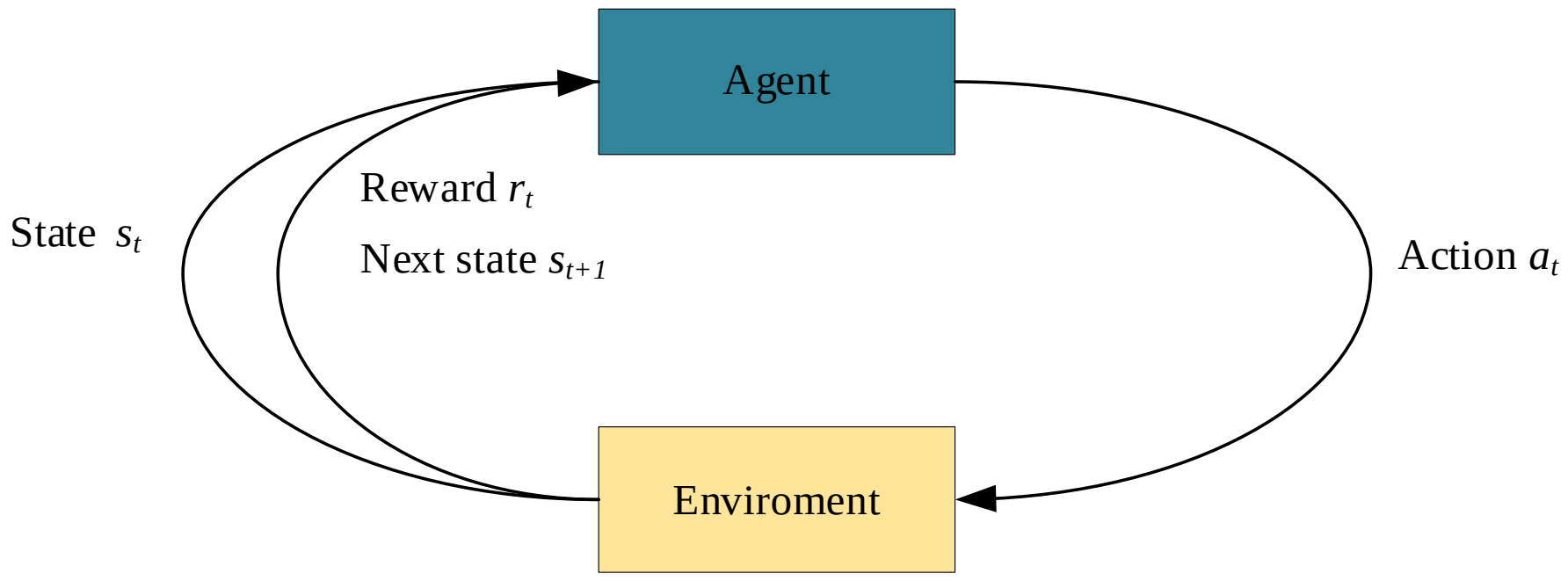

**Fig. 5. Markov decision process**

1) $S$ is the environment state space; $s_t \in S$ refers to the state that the agent perceives in the environment at time $t$.

2) $A$ is the action space; $a_t \in A$ refers to the action selected by an agent at time $t$.

3) $R$ is the reward function; $r_t \in R(s_t, a_t)$ represents the instant reward obtained by the agent choosing action $a_t$ in state $s_t$.

4) $\pi$ is the policy, which represents the mapping relationship from environment state space $S$ to action space $A$.

Fig. 5 shows a typical MDP [40]. At each time step $t$, the agent learns the state $S$ at this moment, chooses an action $a$ according to the current strategy $\pi$, enters the next state, and obtains an instant reward $r$ for the action $a$.

In general, all these algorithms can be divided into two categories: value-based and policy-based algorithms.

The DPPO algorithm used in this study is based on the proximal policy optimization (PPO) algorithm; a distributed design is added, which is an actor-critic framework algorithm to address decision-making problems in a continuous space [40]. The PPO algorithm includes three networks, namely two actor networks and one critic network. This critic network is used to guide the updating direction of the actor networks, and the difference between the two actor networks is used to limit the updating range. To obtain good performance in scenes with multiple uncertainties, the agent must learn as many different experiences as possible. Therefore, a distributed design similar to that in reference [41] is introduced to the PPO algorithm. Multiple agents collect data in different threads simultaneously, and all parallel agents share a common global agent. The relationship between the return function $R_\pi$ and value function $V_\pi$ is mathematically defined in equations (33)-(34). The return function $R_\pi$ refers to all discount rewards from time $t$, and the value function $V_\pi$ refers to the expected reward of the agent acting according to policy $\pi$ from time $t$:

$$V_\pi(s_t, a_t) = \sum_{a_t \in A} \pi(a_t \mid s_t) R_\pi(s_t, a_t), \tag{33}$$

where $\pi(a \mid s)$ represents the probability of choosing action $a$ in state $s$;

$$R_\pi(s_t,\ a_t) = r(s_t, a_t) + \gamma \sum_{s_{t+1} \in S} V_\pi(s_{t+1}), \tag{34}$$

where $\gamma$ is the discount coefficient, which reflects the ratio of future rewards in the present time. Therefore, the advantage function $A_\pi(s, a)$ can be calculated using the following formula:

$$A_\pi(s,a) = R_\pi(s,a) - V_\pi(s,a). \tag{35}$$

The DPPO algorithm revises the strategy according to the size of $A_\pi(s,a)$. The larger $A_\pi(s,a)$ is, the more the policy is updated. However, if the updated degree of the strategy is too large, the algorithm does not converge easily, so the change ratio $r = \frac{\pi(a_t \mid s_t)}{\pi_{old}(a_t \mid s_t)}$ is introduced to reflect the degree of difference between the new and old strategies, and the ratio is limited to a certain range. This method does not only ensure that the strategy is constantly updated in the right direction but also solves the problem of convergence. Further details about the update of the PPO algorithm can be found in [40].

**4.2 Scheduling framework based on deep reinforcement learning**

This study uses DRL to solve the optimal scheduling problem of island IESs, assuming multi-uncertainties from power sources and loads. Before application, the mathematical expressions of the dynamic economic dispatch model of island IESs in section 3 must be transformed into the framework of deep reinforcement learning. In this study, the island IES is the environment for the agent to explore and operate, and the agents adjust the outputs of the equipment in the system for optimal scheduling.

For this island IES, the dynamic environmental states of the system include the user's electrical load demand, heat load demand, freshwater demand, and renewable energy output. Therefore, the dynamic environment state space $S$ include:

$$S = [p_{load,t}, h_{load,t}, w_{load,t}, p_{wt,t}], \tag{36}$$

In a certain time period, the actions that the IES can choose are the outputs of the devices, including the electric power and thermal power of the CHP unit, the electric power and water production rate of the CWP unit, the electric power of the GT, and the thermal power of the GB. Therefore, the action space $A$ include:

$$A = [p_{chp,t}, h_{chp,t}, p_{cwp,t}, w_{cwp,t}, p_{gt,t}, h_{gb,t}], \tag{37}$$

To minimize the operating cost of island IESs, the normalized operating cost is taken as the reward function to guide the agents to explore. Simultaneously, to accelerate the algorithm convergence, the unbalanced power between the source and the load is added to the reward function as a penalty term, so the reward function in this model is expressed as follows:

$$R = -C_E - D + U, \tag{38}$$

where $U$ denotes a suitable positive number that avoids that the reward is always negative; $D$ refers to the unbalanced power between the power sources and the loads, which can be described by the following formula:

$$\begin{aligned} D = &\left| p_{chp,t} + p_{cwp,t} + p_{gt,t} + p_{wt,t} - p_{load,t} \right| + \\ &\left| h_{chp,t} + h_{gb,t} - h_{load,t} \right| + Q\left| w_{cwp,t} - w_{load,t} \right| \end{aligned}. \tag{39}$$

### 4.3 Flowchart of the proposed method

A flowchart of the proposed method is presented in Fig. 6. The solution process can be summarized as follows:

Step 1: Model an island IES, comprehensively considering the use of renewable energy and multiple forms of resource demand.

Step 2: Construct a scheduling model of the island IES with the goal of minimizing operating costs.

Step 3: Transform the scheduling model of the island IES into a DRL framework.

Step 4: Set the hyper-parameters and neural network structure of the DPPO algorithm.

Step 5: Generate a large number of random scenes based on historical data and use them as the training environment.

Step 6: Let agents interact with the environment to train and collect data.

Step 7: The critic network uses the data to calculate the loss function and completes its own update; the actor network uses the data to calculate the change ratio and clip it within a certain range.

Step 8: Use the clipped change ratio and loss function obtained by the critic network to calculate the loss function of the actor network.

Step 9: Update the actor network according to its loss function.

Step 10: Determine whether the algorithm converges to the maximum reward. If this is the case, the decision network is saved; otherwise, go to step 6 and increase the number of training iterations.

Step 11: Apply the saved decision-making network to the dispatch of the island IES.

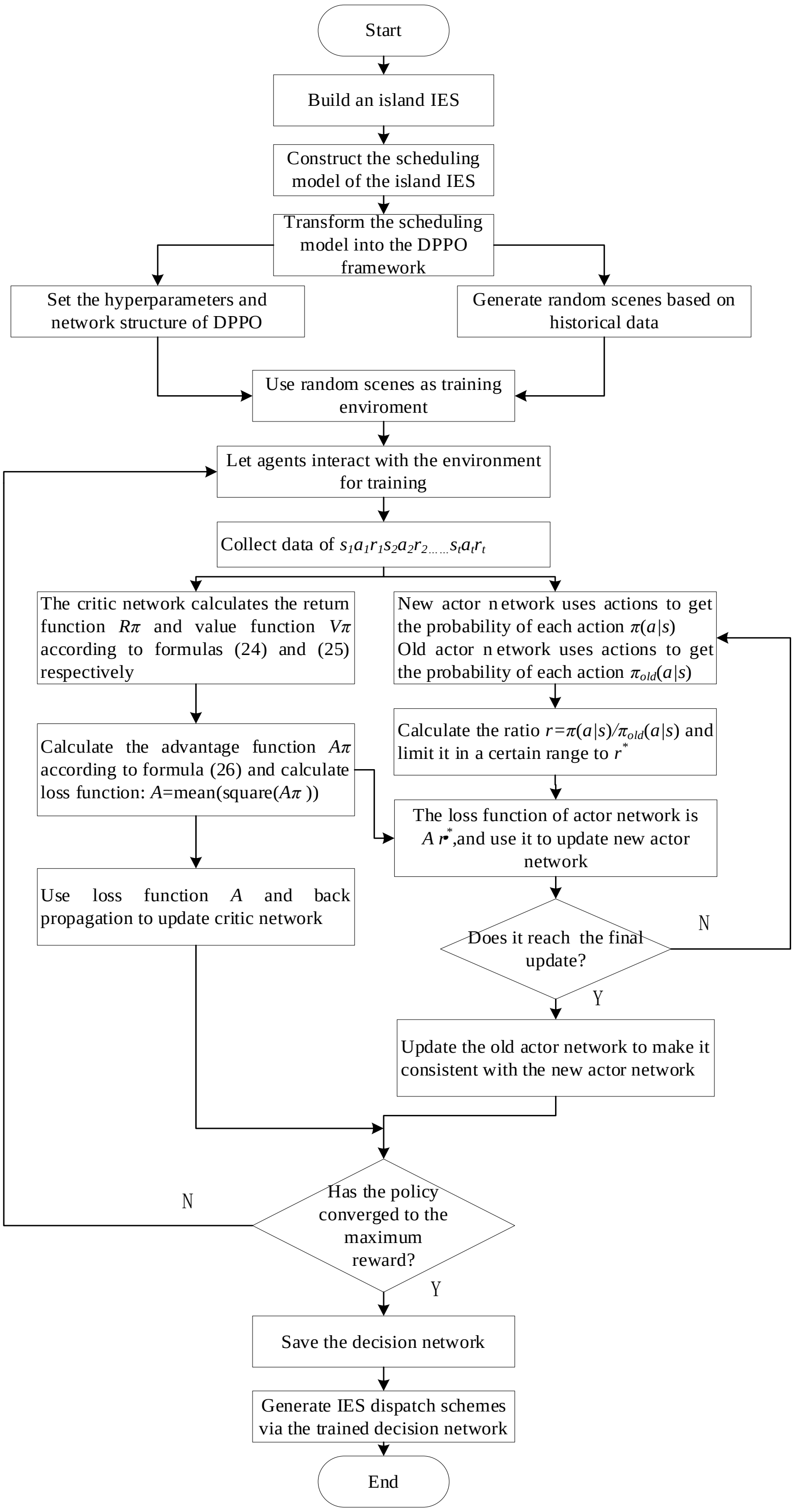

**Fig. 6. Flowchart of the proposed method**

## 5 Case study

To prove the superiority of the proposed approach, the test model shown in Fig. 1 was constructed. The scheduling period of this island IES is taken as 24h [42], and the operating parameters of the main devices are listed in Table 2. The DPPO algorithm uses Python3.6 as the programming language and was implemented with TensorFlow. All simulations and tests were performed on a computer equipped with a dual-core Intel CPU and 16-GB RAM.

**Table 2. Operating parameters of the main devices.**

| Device | Parameter | Symbol | Value |
|---|---|---|---|
| CHP | power upper limit | $p_{chp\max}$ | 5000 kW |
| | power lower limit | $p_{chp\min}$ | 1000 kW |
| | thermoelectric ratio | $b$ | [ 0, 1.4] |
| | ramp rate limit | $RP_{chp\max}$ | 3500 kW/h |
| CWP | power upper limit | $P_{tp\max}$ | 5000 kW |
| | Hydroelectric conversion coefficient | $Q$ | 8 |
| | ramp rate limit | $RP_{tp\max}$ | 3500 kW/h |
| | volume of freshwater | $V$ | 500 m$^3$ |
| | pressure coefficients | $\lambda$ | 4.17 (MPa • L)/mol |
| | initial seawater concentration | $\vartheta_0$ | 0.6 mol/L |
| GT | power upper limit | $p_{gt\max}$ | 3000 kW |
| GB | power upper limit | $h_{gb\max}$ | 3000 kW |

### 5.1 Algorithm training

Before applying the DPPO algorithm to the dispatch of the IES, it must be trained. To improve the ability to deal with multi-uncertainties, a large number of random scenes were generated as the training environment. These scenes were based on the historical data of the resource demand and outputs of WTs from a certain island in northern China [R43]. The hyper-parameters were selected according to [44]. The hidden layers of the actor and critic networks in the DPPO algorithm comprise both two layers, the first with 300 neurons, and the second with 100 neurons. The discount factor was set to 0.95, the mini-batch size was set to 128, the actor's learning rate was set to 0.00005, the critic's learning rate was set to 0.0002, and the number of training episodes was 150000. The average reward curve during training is shown in Fig.7.

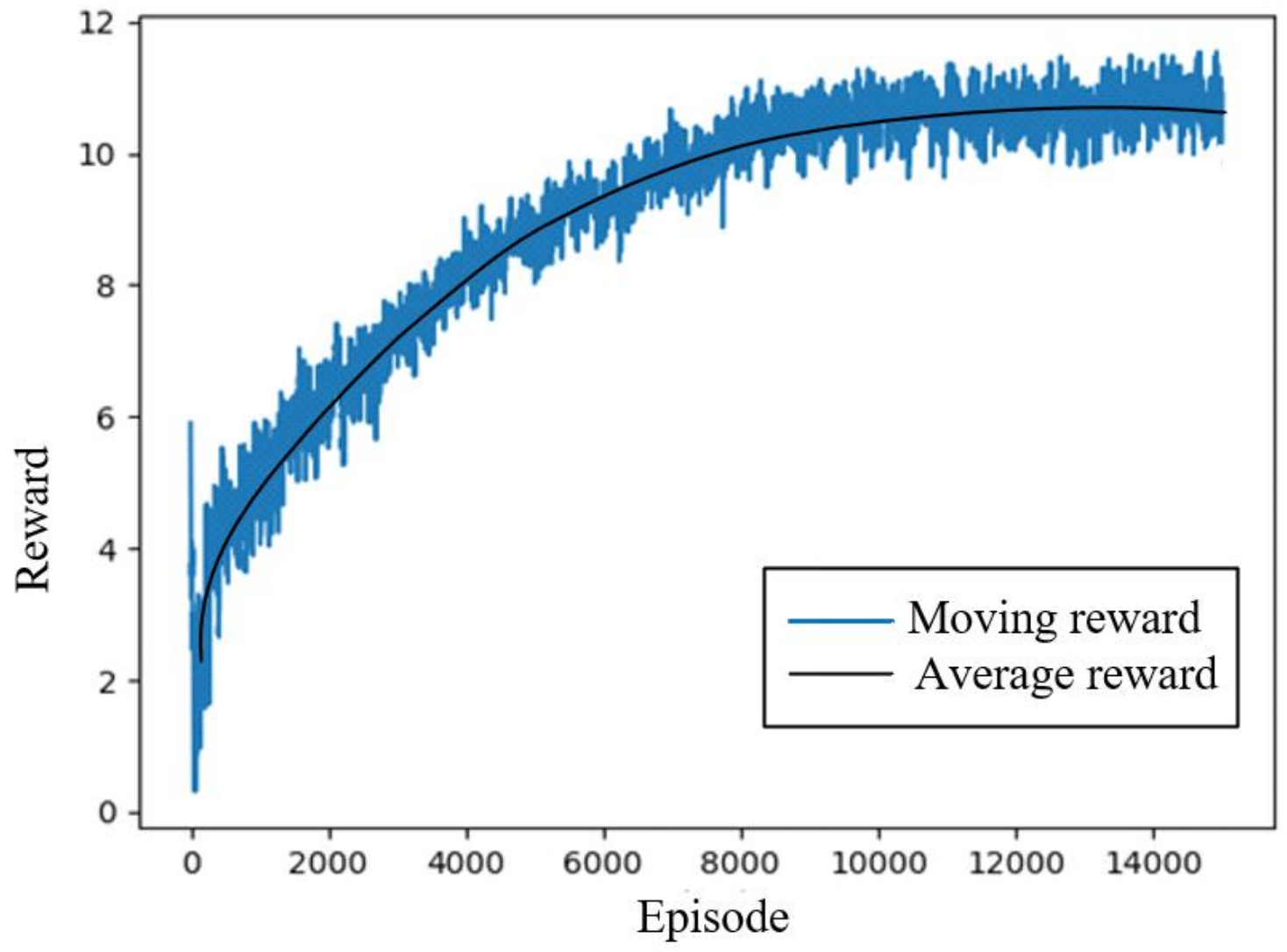

**Fig. 7. Average reward curve during training**

Fig. 7 shows that the algorithm converged after approximately 10,000 episodes, thereby obtaining the optimal dynamic economic dispatch strategy network. Note that, given that the agent is initially unfamiliar with the environment, the agent receives a small reward after executing the scheduling decision. As the number of training episodes increases, the agent continuously interacts with the environment and gains experience, so the reward value shows an overall upward trend and eventually converges. This means that the agent has learned a scheduling strategy that minimizes the operating cost. Because the initial state of each training episode is different, the rising process of the reward during the training process fluctuates.

## 5.2 Analysis of scheduling results based on DPPO

After training, the obtained decision-making network was saved for the economic dispatch of the island IES. To prove the validity of the proposed scheduling model, data from a typical day's resource demand and renewable energy output were selected for testing. The combination of equipment outputs in each time period obtained from the decision-making network is shown in the following figures: Fig. 8 illustrates the electric power dispatch subgraph, Fig. 9 is the thermal power dispatch subgraph, and Fig. 10 shows the freshwater dispatch subgraph.

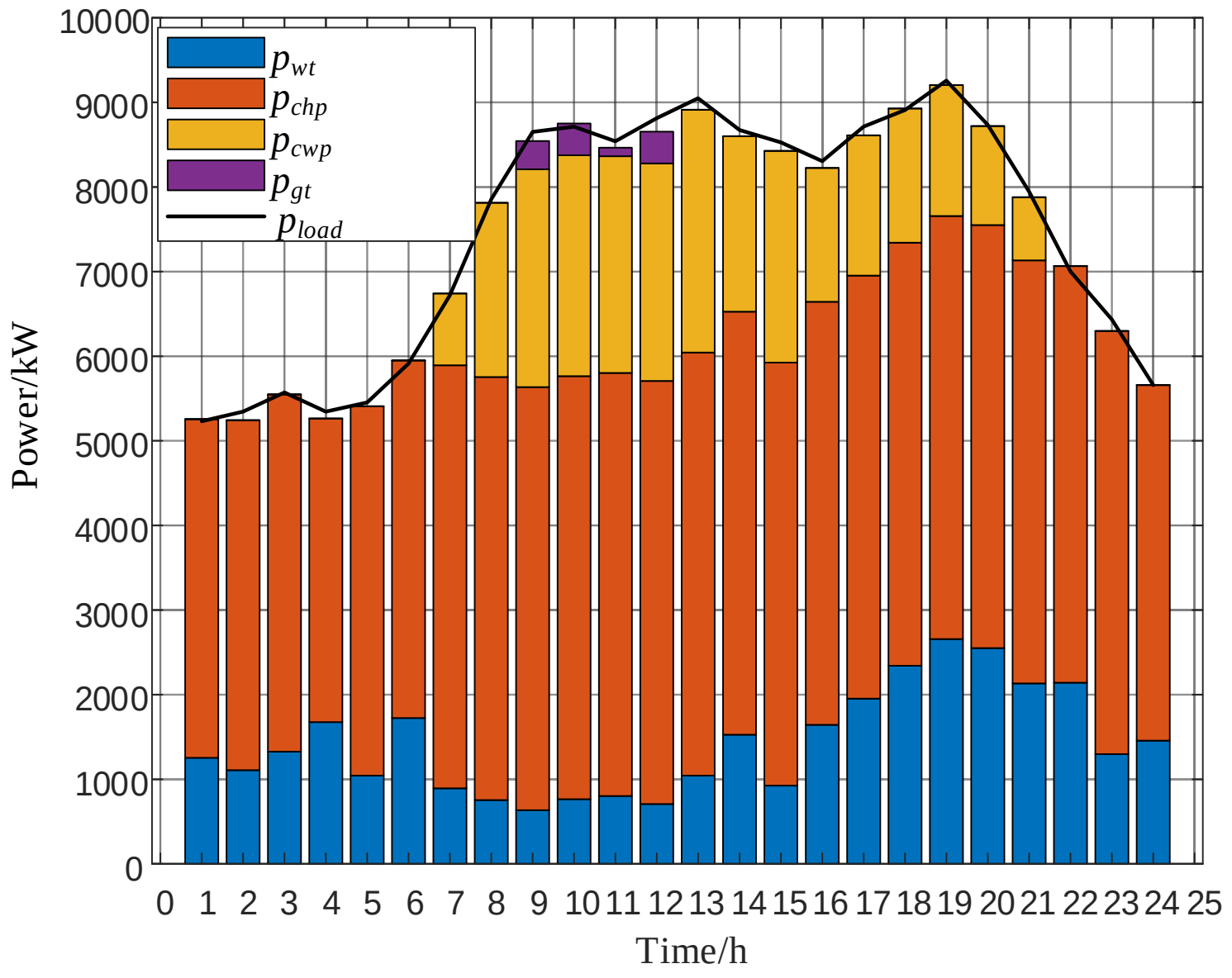

**Fig. 8. Electric power dispatch subgraph**

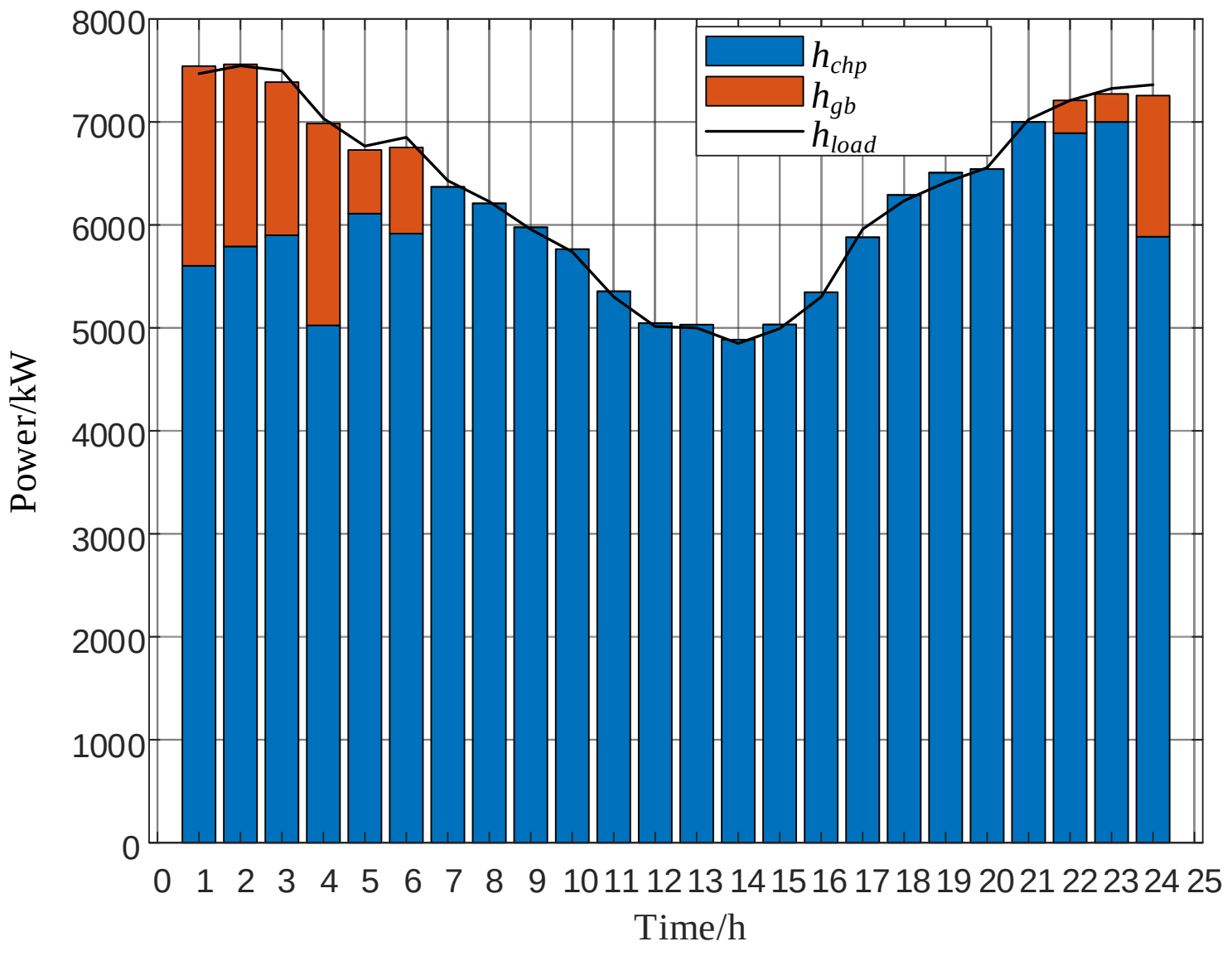

**Fig. 9. Thermal power dispatch subgraph**

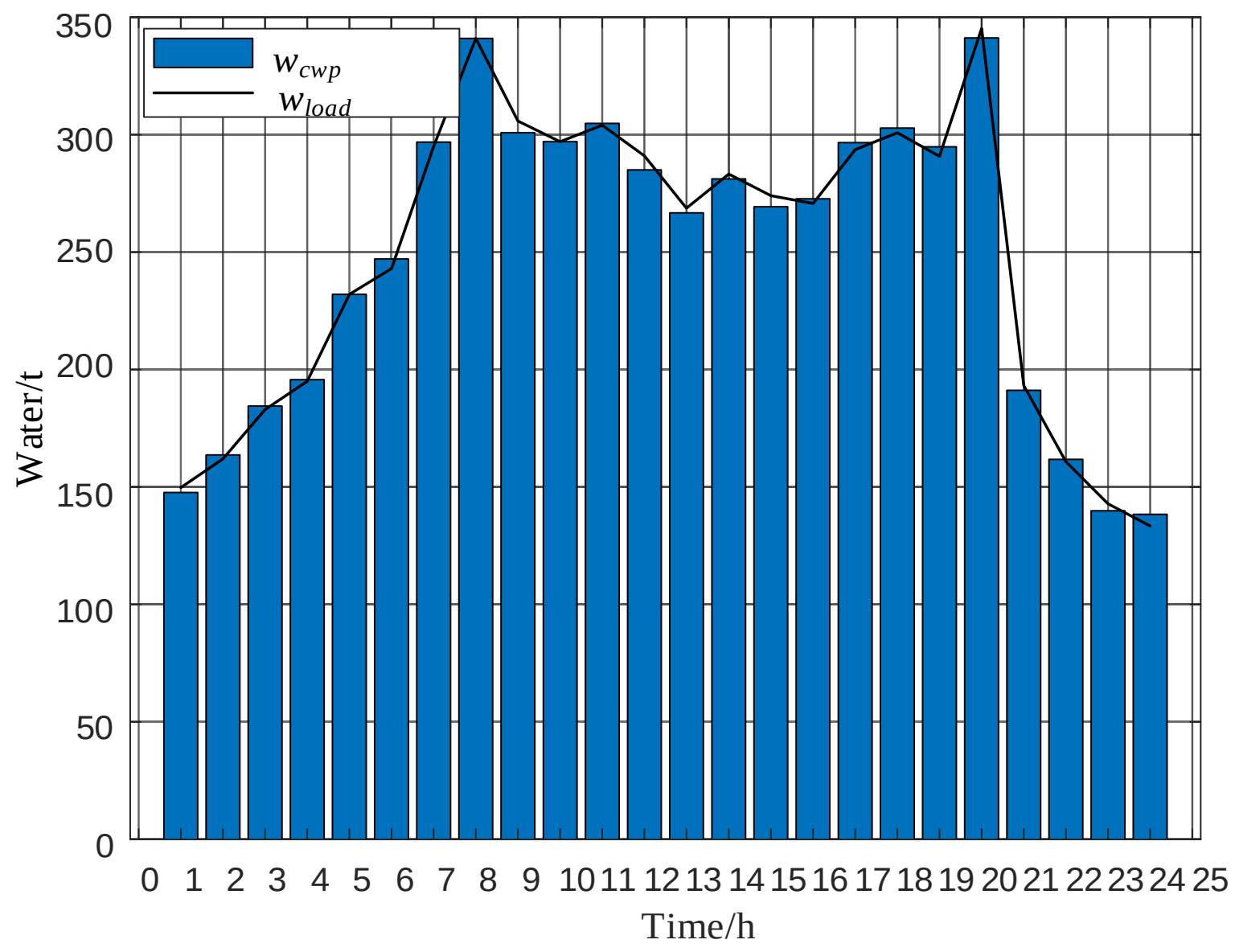

**Fig. 10. Freshwater dispatch subgraph**

Note from the above three figures that the output combination of each equipment made by the decision-making network can properly meet the resource demand of the current dispatch period. The unbalanced power between the source and the load only appears in a few time periods, and it is almost negligible. According to Fig. 8, the output of the WT has the highest priority to achieve complete consumption of renewable energy output; as the load increases, the CHP unit, CWP unit, and GT are put into operation sequentially. This is because the unit operating cost of the co-generation unit is lower, so the co-generation unit is preferentially put into operation. As can be seen from Fig. 9, for economic consideration, in the daytime period in which the heat load is low, the heat demand is mainly supplied by the CHP unit; when the temperature drops at night and the CHP unit cannot meet the demand of heat load, the GB is used as a supplement. In period 4, the output of the GB is high because the electric power of the CHP unit in this time period is small owing to the electro-thermal coupling relationship, and the heat power of the CHP unit in this period is also small, so there is a large demand for the GB to supply heat. Freshwater is a critical resource on the island, and it can be seen from Fig. 10 that the scheduling scheme can provide a stable supply of freshwater for the island.

### 5.3 Analysis of economy and decision efficiency

To prove the economy and decision efficiency of the proposed scheduling method, it is compared with methods using other DRL algorithm and population-based optimization algorithms that can realize real-time continuous control of IESs. The DRL algorithm used for comparison is the soft actor-critic (SAC) algorithm [45], and the selected population-based optimization algorithms are the particle swarm optimization (PSO) algorithm [46] and whale optimization algorithm (WOA) [47-48]. In addition, we also compared with the model-driven interior point method (IPM), and took the step length α=0.3 and central parameter β=0.9 [49]. Concerning the SAC algorithm, its hyper-parameters and network structure are the same as those of DPPO; in WOA and PSO, the population size was set to 100, and the maximum number of iterations was set to 600. The operating costs of the five algorithms for each scheduling period are shown in Fig. 11. Taking into account that the optimization results of the WOA and PSO algorithms, as heuristic algorithms, are slightly different in each run, their operating costs were recorded by averaging the results of 20 independent runs.

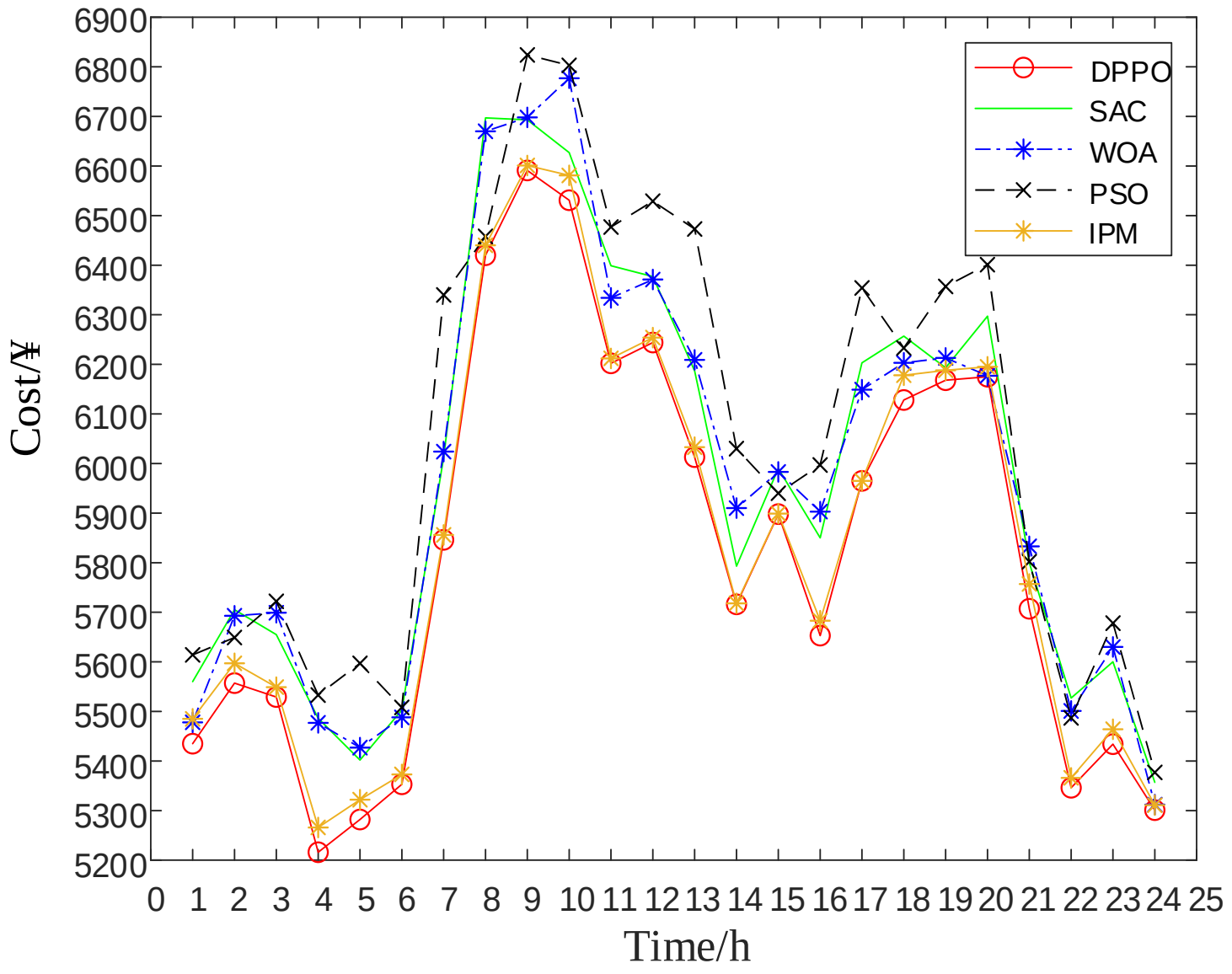

**Fig. 11. Operating costs of the five algorithms**

Fig. 11 shows that the operating cost of our proposed method in almost any time period is lower than that of the other algorithms. The comparison of the daily operating costs and calculation times of different approaches is shown in Table 3. As for the DRL-based approach, we also recorded their training time.

**Table 3. Comparison of economy and decision efficiency for different approaches**

| | Training time /min | Operating cost /¥ | Calculation time /s |
|---|---|---|---|
| Proposed method | 19.6 | 139770.95 | 0.0137 |
| SAC | 51.5 | 143170.82 | 0.0164 |
| WOA | --------- | 143012.27 | 16.459 |
| PSO | --------- | 145169.15 | 55.898 |
| IPM | ---------- | 140458.29 | 29.1364 |

Table 3 shows that the daily operating cost of the proposed method is the lowest, the operating cost of the PSO-based method is the highest, and the operating costs of the WOA-based, SAC-based, and IPM-based methods lie halfway in between. In particular, the operating cost of the proposed method reaches a minimum value of 139970.95 Yuan, which is 5398.24 Yuan below that of the PSO-based method, and 3399.87 and 3241.32 Yuan below those of the SAC-based and WOA-based methods, respectively. Although the performance of IPM-based scheduling method in operating cost is not bad, it relies on accurate modeling, so it is only suitable for day ahead scheduling with fixed schedule and the scheduling decision cannot be adjusted in real time with the change of external environment. It is now evident that the method proposed in this study achieves better optimization results than the other three methods. For the training time, the proposed method is much smaller than the SAC algorithm, which is also DRL-based, indicating that our designed distributed structure can greatly improve the data collection rate and thus accelerate the convergence of the algorithm. In terms of calculation time, the trained DPPO and SAC decision-making networks can obtain scheduling decisions almost instantaneously after the perception of the external environment, which is significantly less than other methods. This is because, during the training process, as deep reinforcement algorithms, DPPO and SAC have explored as many situations as possible, and use deep neural networks to save the acquired strategies. Although the SAC-based method has almost the same decision-making efficiency as the proposed method, its economic performance is still somewhat worse. The proposed method has good decision efficiency and can respond to changes in the external environment more quickly, so it is more suitable for the scheduling of islands IESs with multi-uncertainties.

### 5.4 Analysis of uncertainty handling capacity

To test the ability of the DPPO decision-making network to handle various uncertainties, emergency simulations for typical scenarios were performed. Specifically, the electrical load was suddenly increased during period 1 when the electrical load was low, and the electrical load was suddenly decreased during period 19 when the electrical load was high; in period 9, when the WT output was small, the output of the WT was suddenly increased, in period 20, when the WT output was large, the output of the WT was suddenly decreased to zero; the heat demand was suddenly decreased in period 2 when the heat load was high, and was sharply increased in period 14 when the heat load was small. The strategies to handle the changes in electrical load demand are shown in Fig. 12 and 13; the strategies to manage the changes in the WT output are shown in Fig. 14 and 15; finally, the strategies to handle the changes in heat load demand are shown in Fig. 16 and 17.

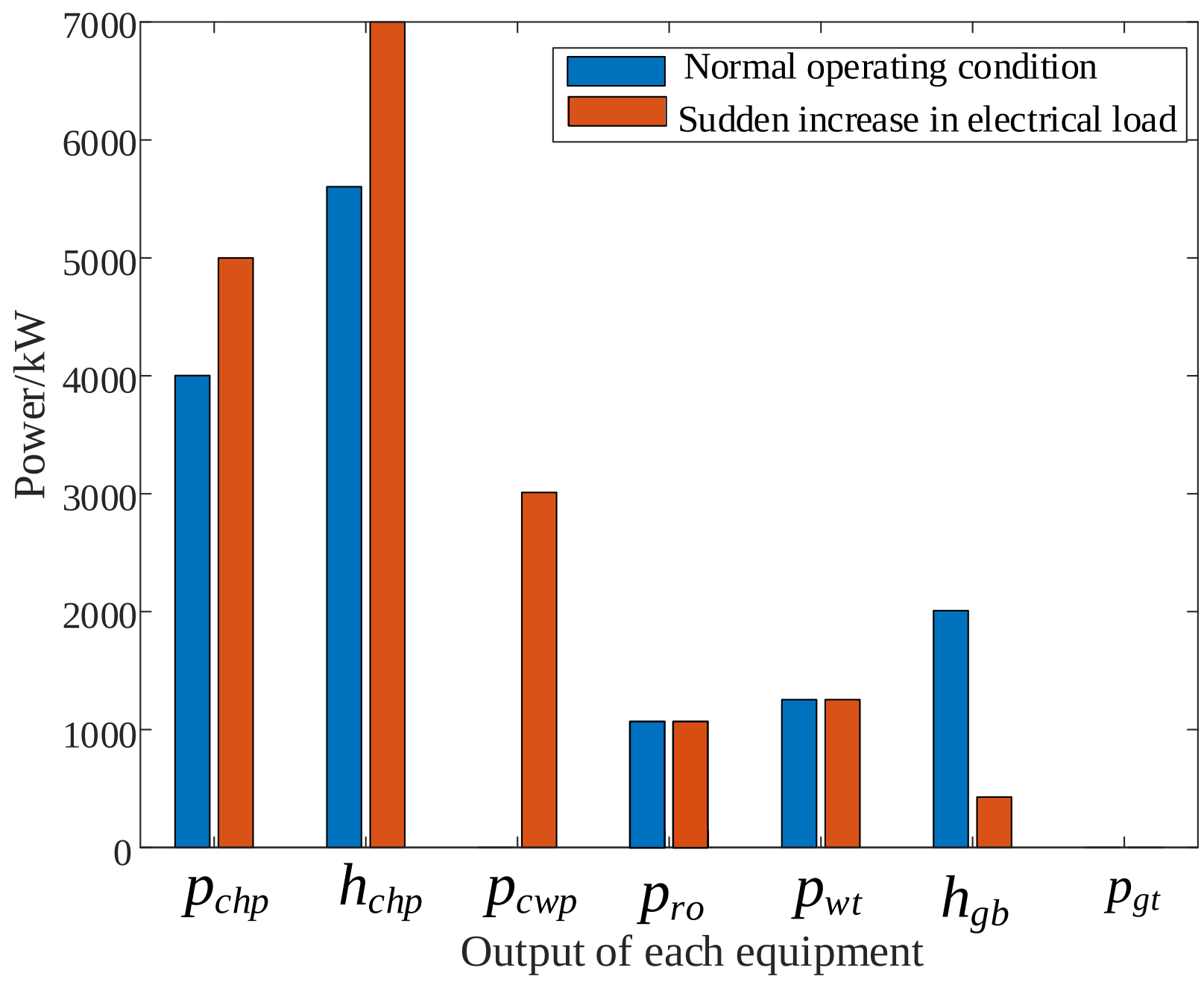

**Fig. 12. Scheduling strategy for sudden increase in electrical load**

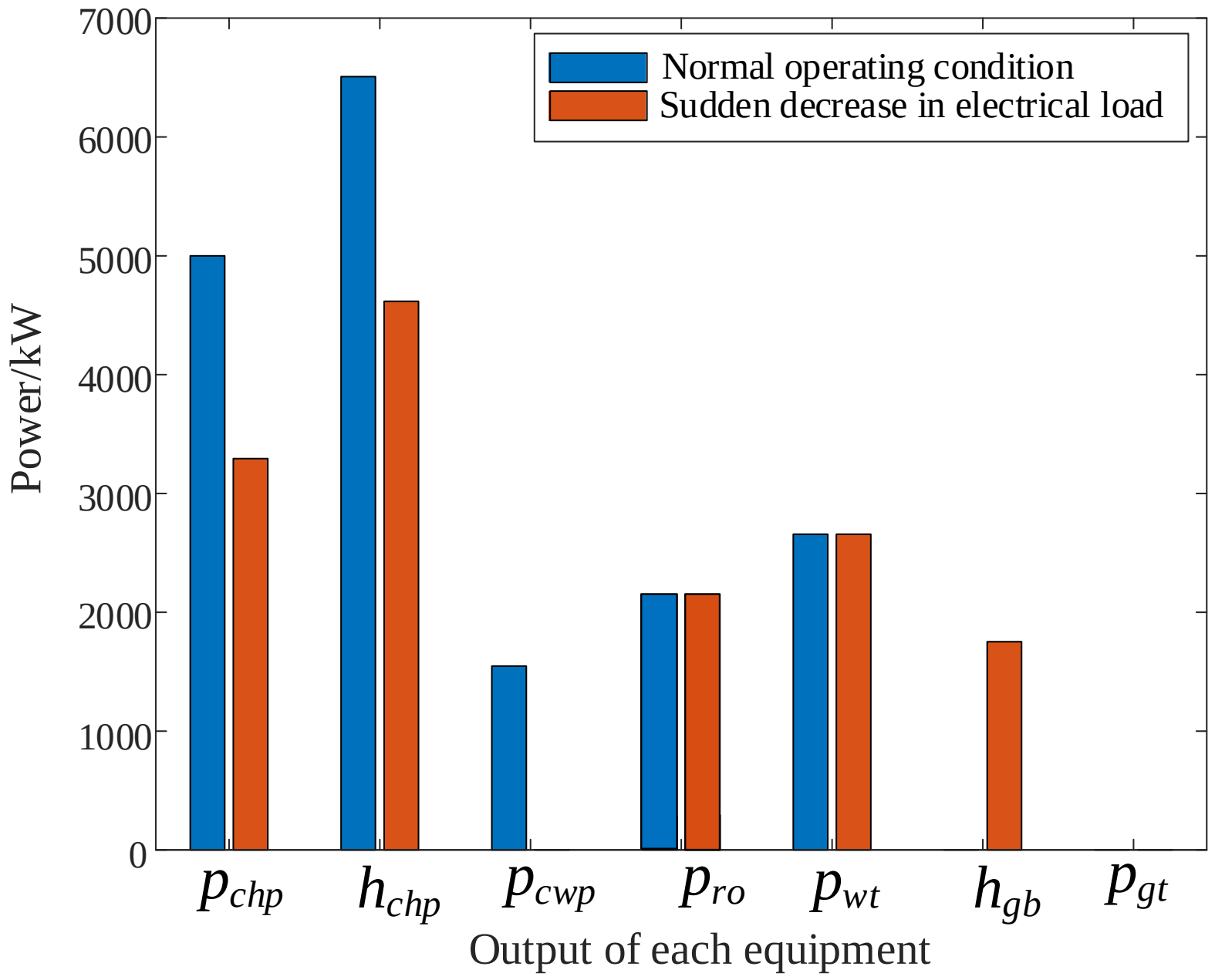

**Fig. 13. Scheduling strategy for sudden decrease in electrical load**

As shown in Fig. 12, when the user's electricity load was suddenly increased, the DPPO decision-making network made the CHP unit and the CWP unit increase the electric power to compensate for the electric power gap. In addition, as the electric power of the CHP unit increased, its thermal power increased as well. Therefore, the thermal power of the GB was reduced. Fig. 13 shows that when the user's electrical load was suddenly decreased, the DPPO decision-making network forced the CHP unit and CWP unit to reduce the electric power, and as the electrical power of the CHP unit decreased, its thermal power also decreased. To meet the user's heat demand, the GB compensated for the thermal load gap.

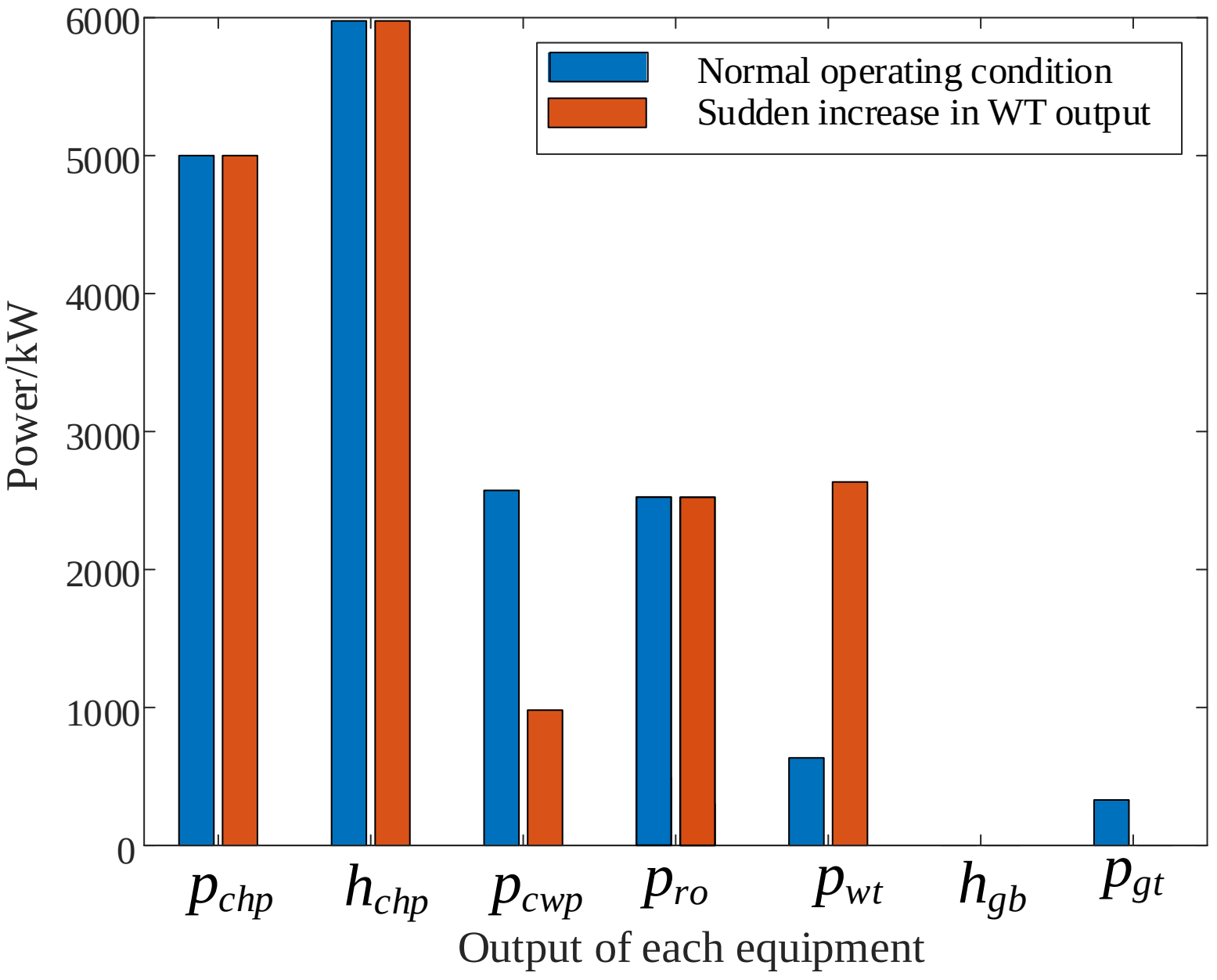

**Fig. 14. Scheduling strategy for sudden increase in WT output**

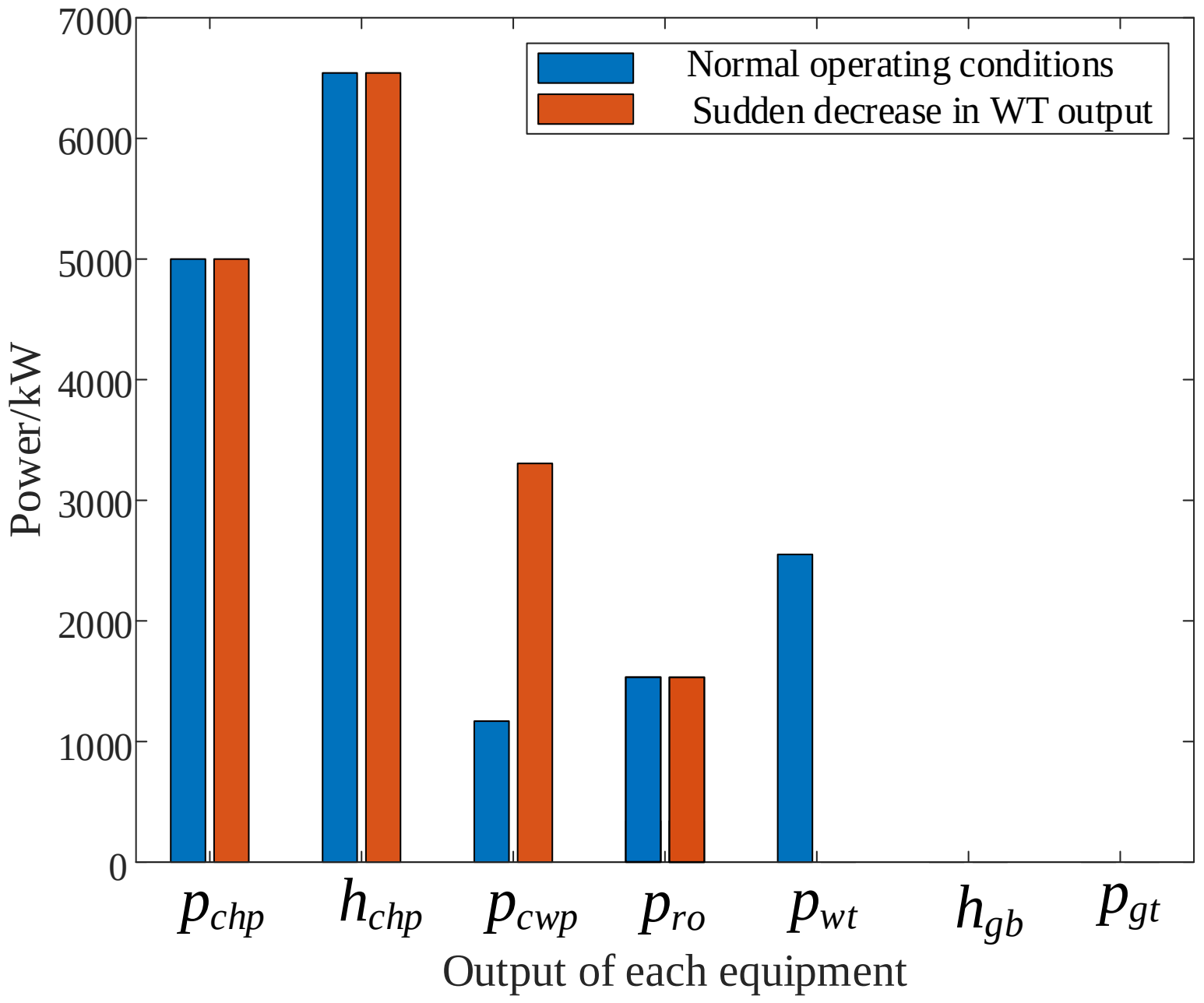

**Fig. 15. Scheduling strategy for sudden decrease in WT output.**

As shown in Fig. 14, when the output of the WT was increased suddenly, the DPPO decision-making network first reduced the power of the GT to completely consume the wind power. Given that it could not achieve the complete consumption of wind power, it reduced the electrical power of the CWP unit. Note from Fig. 15 that when the WT was suddenly decreased and the electric power was set to zero, given that the CHP unit had reached its upper power limit at this time, the DPPO network forced the CWP unit to increase the electric power to compensate for the electric load gap.

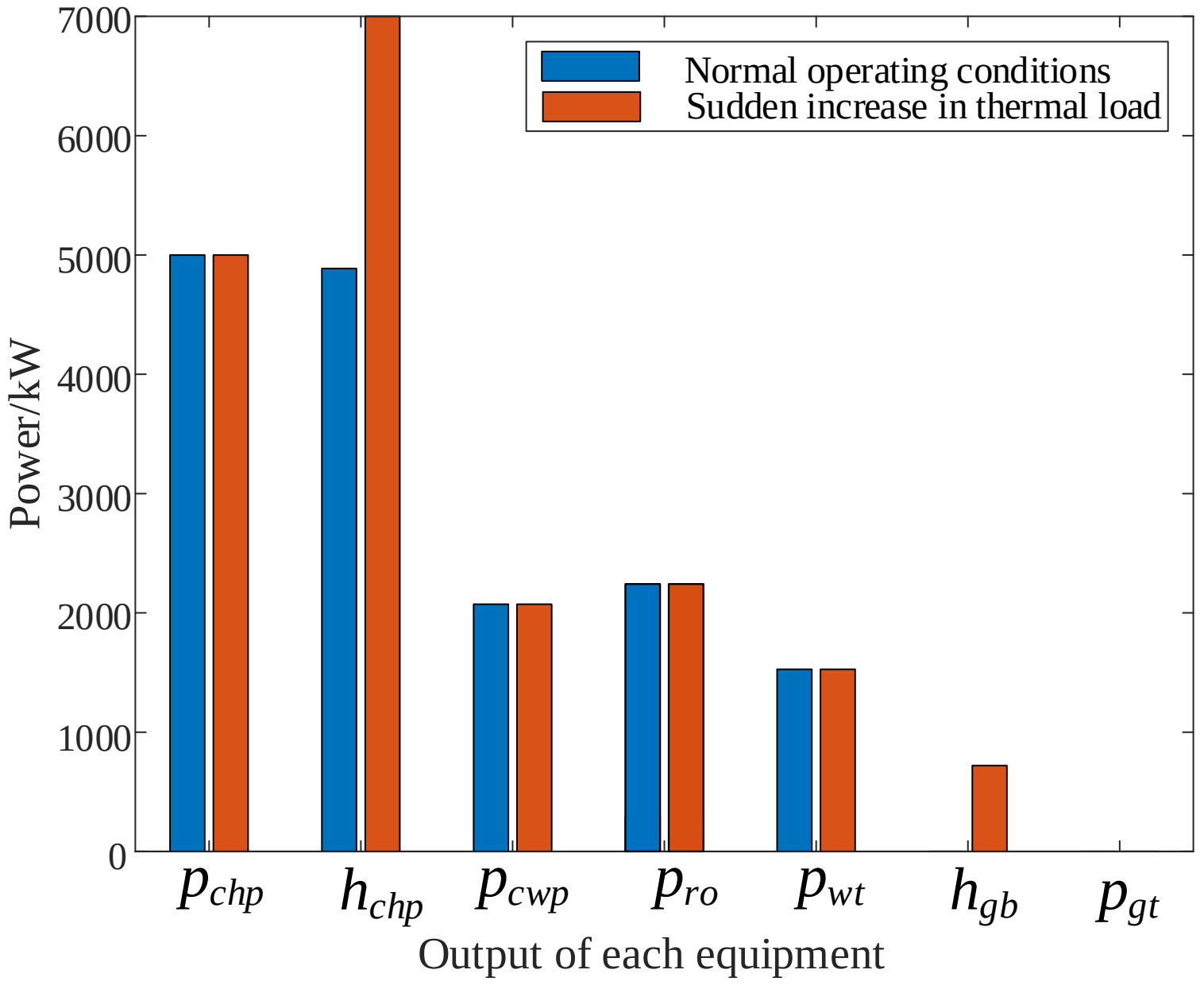

**Fig. 16. Scheduling strategy for sudden increase in thermal loads**

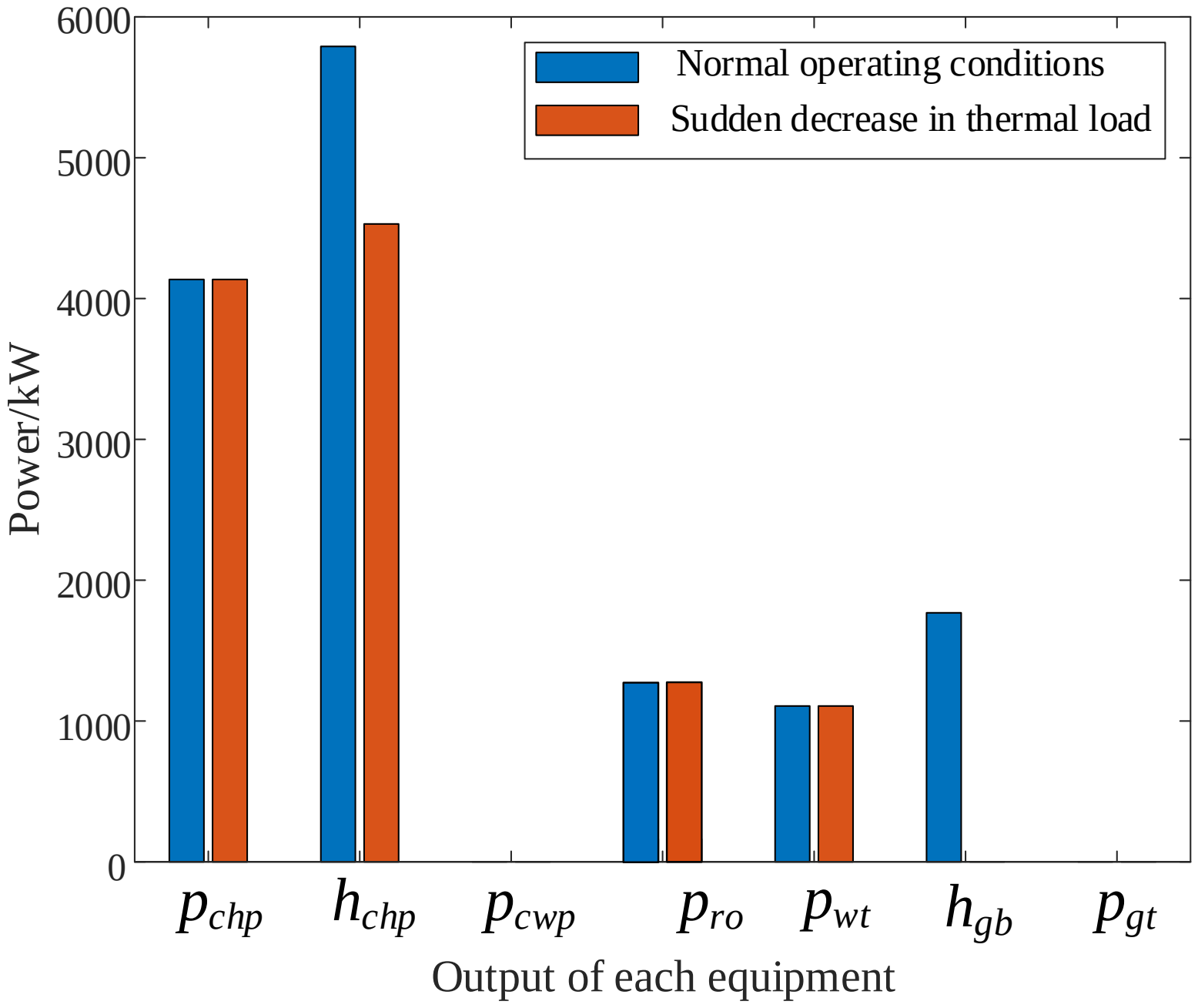

**Fig. 17. Scheduling strategy for sudden decrease in thermal loads**

Fig. 16 shows that when the user's thermal load was suddenly increased, the DPPO decision-making network first increased the thermal power of the CHP unit, and when the CHP unit reached its upper power limit, the GB was used to fill the thermal power gap. Fig. 17 indicates that when the user's thermal load was suddenly decreased, the DPPO decision-making network first reduced the heat power of the GB, and then reduced the thermal power of the CHP unit when the power of the GB was reduced to 0 kW.

According to the testing of the above emergencies, it can be concluded that when the external environment changes, the DPPO network can adjusts the scheduling strategy in real time according to the current environmental status without manual intervention. This indicates that the DPPO decision-making network has a good ability to deal with various uncertainties and emergencies. In addition, it can be seen from the priority of the power rise and fall of each device, this decision-making strategy of the DPPO network is logical. It is worth noting that although we only tested on a typical day, the test results for these extreme cases show that the method proposed in this paper is suitable for a variety of scheduling scenarios.

## 6 Conclusions

To achieve a stable supply adapted to the multiple energy demands in islands, and strengthen the use of renewable energy resources, this study proposes an optimal scheduling approach based on DRL for island IESs considering source-load multi-uncertainties and hydrothermal simultaneous transmission. Based on the above simulation results, the following conclusions can be drawn:

(1) The proposed island IES dispatching method manages to realize a stable supply of multiple energy demands, including electricity, heat, and freshwater, and the full utilization of renewable energy output.

(2) The constructed HST mode can achieve energy and material transmission simultaneously, which can greatly improve the utilization efficiency of freshwater and alleviate the shortage of freshwater on islands. This provides a new reference for thermal energy and water supply in areas with a shortage of freshwater resources.

(3) The DPPO-based dispatching approach is capable of automatically adapting to a variety of uncertain changes by the interaction between agents and the environment, thereby achieving the energy balance of the island IES without the prediction of uncertain renewable power outputs and loads. Furthermore, it accomplishes better optimization results and higher calculation efficiency than other algorithms.

(4) The simulation results based on data from a real island proved the effectiveness of the proposed method. In addition, the proposed method demonstrated a good ability to deal with uncertainties, when an emergency occurs, the dispatch plan can be adjusted without manual intervention in real time.

As a model-free approach, the presented method is applicable to island IESs with volatility on both the source and load sides. This study did not consider the heat loss of freshwater heaters in HST modeling, while a more realistic model should be considered. The next step will focus on how to configure water and heat storage devices with appropriate capacity to break the supply and demand balance constraint of freshwater and improve the flexibility of IES. The handling of uncertainty is the core problem for IES optimal scheduling, and it can be divided into DRL based method and mathematical analytic method depending on whether the uncertainty constraints require deterministic transformation. Scenario-based methods, which have been rapidly developed in recent years, hold great promise for future applications in dealing with uncertainty problems. It would also be interesting to investigate the resilient scheduling under cyber attacks, e.g. false data injection attacks [50].

## ACKNOWLEDGEMENT

This work is supported by the National Natural Science Foundation of China under Grant U2066208 and Grant 62233006, and the Natural Science Foundation of Jilin Province, China under Grant YDZJ202101ZYTS149.